\documentclass[12pt]{article}

\usepackage{amsmath,amssymb,amsfonts,graphics,graphicx,amscd,amsfonts,epsf,epsfig,color}
\newcommand {\beq} {\begin{equation}}
\newcommand {\eeq} {\end{equation}}
\newcommand {\beqa}{\begin{eqnarray}}
\newcommand {\eeqa}{\end{eqnarray}}

\newcommand {\del} {\partial}
\newcommand {\tr}{{\rm tr\,}}

\newcommand {\ee}{\mbox{e}}

\newcommand{\bbR}{{\mathbb R}}
\newcommand{\bbC}{{\mathbb C}}

\makeatletter
    
    \@addtoreset{equation}{section}
  \makeatother

\allowdisplaybreaks[4]

\date{}
\begin{document}

\begin{flushright} 
KEK-TH-1855
\end{flushright} 

\vspace{0.1cm}

\begin{center}
  {\LARGE
Justification of the complex Langevin method
\\
[0.2cm]
with the gauge cooling procedure
%
  }
\end{center}
\vspace{0.1cm}
\vspace{0.1cm}
\begin{center}

         Keitaro N{\sc agata}$^{a}$\footnote
          {
 E-mail address : knagata@post.kek.jp},  
         Jun N{\sc ishimura}$^{ab}$\footnote
          {
 E-mail address : jnishi@post.kek.jp} 
and
         Shinji S{\sc himasaki}$^{a}$\footnote
          {
 E-mail address : simasaki@post.kek.jp} 

\vspace{0.5cm}

$^a${\it KEK Theory Center, 
High Energy Accelerator Research Organization,\\
1-1 Oho, Tsukuba, Ibaraki 305-0801, Japan}

$^b${\it Graduate University for Advanced Studies (SOKENDAI),\\
1-1 Oho, Tsukuba, Ibaraki 305-0801, Japan} 

\end{center}

\vspace{1.5cm}

\begin{center}
  {\bf abstract}
\end{center}

Recently there has been remarkable progress in
the complex Langevin method, which aims at solving the complex action problem
by complexifying the dynamical variables in the original path integral.
In particular, a new technique called the gauge cooling was
introduced and the full QCD simulation at finite density
has been made possible in the high temperature (deconfined) phase or with heavy quarks.
Here we provide a rigorous justification of the complex Langevin method
including the gauge cooling procedure.
We first show 
that
%
the gauge cooling can be formulated
as an extra term in the complex Langevin equation
involving a gauge transformation parameter,
which is chosen appropriately as a function of the
configuration before cooling.
The probability distribution 
of the complexified dynamical variables
is modified by this extra term.
However, this modification is shown {\em not} to affect the
Fokker-Planck equation for the corresponding complex weight
as long as observables are restricted to gauge invariant ones.
Thus we demonstrate explicitly that the gauge cooling 
can be used as a viable technique
to satisfy the convergence conditions for the complex Langevin method.
We also discuss the ``gauge cooling'' in 0-dimensional systems
such as vector models or matrix models.
%
%
%

\newpage

\section{Introduction}

Monte Carlo calculation has been playing an important role in
nonperturbative studies of quantum field theories.
However, the usefulness becomes quite limited when the action $S$
becomes complex because the integrand $\ee^{-S}$ in the path 
integral can no longer be regarded as the Boltzmann weight.
This occurs in many interesting cases such as
QCD at finite density or with a theta term,
gauge theories with a Chern-Simons term, chiral gauge theories,
and so on. It also occurs in supersymmetric gauge theories
and matrix models relevant to nonperturbative studies
of superstring theory.

Amongst various approaches to this complex action problem,
the one based on the complex Langevin equation has been 
attracting a lot of attention recently.
The original idea was proposed by Parisi \cite{Parisi:1984cs}
and Klauder \cite{Klauder:1983sp}
in 1983, and since then it was applied to various systems with 
complex actions. A salient feature of the method is that
it works beautifully in some fairly nontrivial cases, 
but it fails completely in the other cases. 
For a long time, theoretical understanding of this 
feature was missing, and that led to the gradual decline of 
interest in this approach.
However, in 2011 one of the problems of the method in the case
it fails has been clearly identified \cite{Aarts:2009uq,Aarts:2011ax}.
The authors first derived the key
relation
between the complex Langevin process and the Fokker-Planck
equation for the complex weight.
Then it was found that
the integration by parts used in the derivation
may not be justified
unless the probability distribution of
the complexified dynamical variables 
is suppressed strongly enough when they take large values.

The gauge cooling has been proposed to cure this problem
in the case of gauge theories \cite{Seiler:2012wz}.
It
has been applied to finite density QCD
in the heavy dense limit and shown to work
in the whole parameter regime in that limit \cite{Aarts:2014bwa}.
More recently it has been applied to finite density QCD
without taking the heavy dense limit, and it
is shown to work at least in 
the deconfined phase \cite{Sexty:2013ica}.
These are already quite remarkable since the cases that
have been studied include
a parameter region, which would be hardly accessible
by other methods such as reweighting.
On the other hand, it is also realized in 
a solvable gauge theory with a complex coupling constant
that
there exists some parameter regime in which
the gauge cooling cannot cure completely
the insufficient fall-off of the 
probability distribution \cite{Makino:2015ooa}.
%

In fact there is another problem anticipated to
occur when one applies the complex Langevin method (CLM) to QCD 
with light quarks at low temperature.
This was realized in ref.~\cite{Mollgaard:2013qra}
by applying the CLM
to the Random Matrix Theory
for finite density QCD.
It turned out that a naive implementation of the 
method fails as the quark mass is 
decreased (See ref.~\cite{Mollgaard:2014mga}, however.).
The reason for this failure 
was speculated
to have something to do with the logarithmic singularity 
in the action due to the fermion 
determinant \cite{Mollgaard:2013qra,Mollgaard:2014mga,Greensite:2014cxa}.
On the other hand, it 
was also pointed out \cite{Seiler} that
the problem occurs due to a singular drift term, which
breaks the requirement of holomorphy in the
derivation of the key relation 
between the complex Langevin process and the Fokker-Planck
equation for the complex weight \cite{Aarts:2009uq,Aarts:2011ax}.
Recently two of the authors (J.N.~and S.S.) \cite{Nishimura:2015pba}
have argued that it is actually 
the integration by parts used in the derivation 
that is invalidated by
the singular drift term.
According to this understanding, the problem 
can be avoided if the probability distribution is
suppressed strongly enough
near the singularity.
In a separate paper \cite{NNS} we show that
this 
can also be achieved by the gauge cooling 
if one chooses appropriately the quantity 
that should be reduced by the cooling procedure.

While intuitive arguments for justification of the gauge cooling
are given in the literature (See, for instance, 
section 5 of ref.~\cite{Aarts:2013uxa}),
a rigorous justification has been missing.
Actually there is even some suspicion in the community 
that the procedure may not be fully justified.
Some of the concerns we have encountered 
in private communications are: 
1) the gauge cooling uses a complexified gauge symmetry, 
which is not a symmetry of the original system. 
2) the noise term in the complex Langevin equation is
invariant under the original gauge transformation but
not under the complexified gauge transformation.
3) the quantity one tries to reduce by the 
complexified gauge transformation is not holomorphic, 
which may spoil the justification of the CLM.
In view of this situation,
%
here we provide
a rigorous justification 
of the CLM including the gauge cooling procedure.
%
We first show that 
the gauge cooling can be formulated
as an extra term in the complex Langevin equation
involving a gauge transformation parameter,
which is chosen appropriately as a function of the
configuration before cooling.
The probability distribution of the complexified dynamical variables
is modified by this extra term.
However, this modification is shown not to change
the Fokker-Planck equation for the 
corresponding complex weight
as long as the observables are restricted to
gauge-invariant ones.
Thus we conclude that the gauge cooling can be used to realize the
properties of the probability distribution that are required
for its relation to the complex weight
without affecting
the Fokker-Planck equation.

We also discuss the ``gauge cooling'' in 0-dimensional systems such 
as vector models or matrix models,
which is simpler than in lattice gauge theory.
Apart from pedagogical purposes,
we 
consider that it is useful, for instance,
in studying the matrix models relevant to 
superstring theory \cite{Kim:2011cr,Anagnostopoulos:2013xga}.

The rest of this paper is organized as follows.
In section \ref{sec:ov_gravity}
we briefly review the Langevin method
starting from the well-established case of real action,
and discuss
the conditions for correct convergence
in the case of complex action.
In section \ref{sec:cooling} we discuss
the ``gauge cooling'' in 0-dimensional systems 
and provide its justification.
In section \ref{sec:lattice}
we discuss the application of
the CLM
to lattice gauge theory. 
In section \ref{sec:cooling-lgt}
we present an explicit justification of the gauge cooling
in lattice gauge theory.
Section \ref{sec:conclusion}
is devoted to a summary and discussions.

\section{Brief review of the Langevin method}
\label{sec:ov_gravity}

In this section we briefly review the Langevin method.
(As a comprehensive review on this subject, 
we recommend ref.~\cite{Damgaard:1987rr}.) 
Here we consider a system of $n$ real variables $x_k$ 
($k=1,\cdots ,n$) given by the partition function
\begin{alignat}{3}
Z = \int dx \, \ee^{-S(x)} =  \int \prod_{k} dx_k \, \ee^{-S(x)} \ , 
  \label{eq:part-fn}
\end{alignat}
where the action $S(x)$ is a function of $x=(x_1 , \cdots , x_n)$.
We start with the well-established case of real action, which is 
known also under the name of stochastic quantization.
Then we discuss the case of complex action focusing on the 
conditions for correct convergence.

\subsection{the case of real action}
When the action $S(x)$ is real, we can use 
the ordinary Langevin method to study this system \cite{Parisi:1980ys}.
Introducing a fictitious time $t$, we consider the $t$-evolution
governed by the Langevin equation 
\begin{alignat}{3}
\dot{x}_k^{(\eta)} (t) = - \frac{\del S}{\del x_k} + \eta_k(t) \ , 
\label{eq:Langevin}
\end{alignat}
where $\eta_k(t)$ are probabilistic variables obeying the 
probability distribution 
$\ee^{-\frac{1}{4} \int dt \, \eta_k(t)^2}$.
The first term and the second term on the right-hand side of
the Langevin equation (\ref{eq:Langevin}) are commonly called
the drift term and the noise term, respectively, for historical reasons.

The probability distribution of $x^{(\eta)}(t)$ can be
defined as
\begin{alignat}{3}
P(x,t) = \Bigl\langle \prod_k \delta \Big(x_k - x_k^{(\eta)} (t) \Big)
\Bigr \rangle_\eta \ ,
\label{def-P}
\end{alignat}
where the expectation value $\langle \ \cdots \ \rangle_{\eta}$
is defined by
\begin{alignat}{3}
\langle \ \cdots \ \rangle_{\eta}
= \frac{\int {\cal D}\eta \cdots \ee^{-\frac{1}{4} \int dt \, \eta_k(t)^2}}
{\int {\cal D}\eta  \, \ee^{-\frac{1}{4} \int dt \, \eta_k(t)^2}} \ .
\label{def-EV-eta}
\end{alignat}
Using this notation, one obtains, for instance,
\begin{alignat}{3}
\Big\langle \eta_k(t_1) \, \eta_l (t_2)  \Big\rangle_{\eta}
 = 2 \, \delta_{kl} \, \delta(t_1 - t_2) \ .
\label{2pt-corr-eta}
\end{alignat}
One can actually show that $P(x,t)$ satisfies
the Fokker-Planck (FP) equation 
(See section \ref{sec:discretized-Langevin} for derivation.)
\begin{alignat}{3}
\frac{\del P}{\del t}
= \frac{\del}{\del x_k} 
\left(\frac{\del S}{\del x_k} + \frac{\del}{\del x_k} \right) P \ ,
\label{FPeq}
\end{alignat}
which has a time-independent solution
\begin{alignat}{3}
P_{\rm time-indep}(x) = \frac{1}{Z} \, \ee^{-S(x)} \ .
\label{time-indep-sol}
\end{alignat}

Under quite general conditions \cite{Damgaard:1987rr},
one can show that the eigenvalues
of the differential operator acting on $P$ on the right-hand side
of (\ref{FPeq}) are strictly negative except for
the zero eigenvalue corresponding 
to (\ref{time-indep-sol}). 
This implies that the probability distribution
$P(x,t)$ approaches (\ref{time-indep-sol}) exponentially.
One can therefore obtain a vacuum expectation value (VEV)
with respect to the 
partition function (\ref{eq:part-fn}) as
\begin{alignat}{3}
\Big\langle {\cal O}(x) \Big\rangle 
&= \int 
dx
\, {\cal O}(x) \, P_{\rm time-indep}(x) \nonumber \\
&= \lim_{t \rightarrow \infty} \int 
dx
\, {\cal O}(x) \, P(x,t) 
\nonumber \\
&= \lim_{t \rightarrow \infty} 
\Big\langle  {\cal O} \Big(x^{(\eta)} (t) \Big)
\Big\rangle_{\eta} \nonumber  \\
&= \lim_{T \rightarrow \infty} \frac{1}{T}
\int_{t_0}^{t_0+T} dt \, {\cal O}\Big(x^{(\eta)} (t)\Big) \ .
\label{O-time-av}
\end{alignat}
In the last step,
the statistical average over $\eta$
is replaced by the time average
assuming the ergodicity of the 
stochastic process as is done in usual Monte Carlo methods.

\subsection{the discretized Langevin equation}
\label{sec:discretized-Langevin}

When one tries to solve
the Langevin equation (\ref{eq:Langevin})
numerically,
one has to discretize
the fictitious time $t$ and solve, 
for instance,\footnote{There are more sophisticated
ways for discretization that can be used to reduce the
systematic errors due to the 
discretization. See ref.~\cite{Fukugita:1986tg} and references therein.}
\begin{alignat}{3}
x_k^{(\eta)} (t+\epsilon) 
= x_k^{(\eta)} (t) + \epsilon 
\left(- \frac{\del S}{\del x_k} + \eta_k(t) \right) \ ,
\label{eq:Langevin-discretized}
\end{alignat}
where the probabilistic variables $\eta_k(t)$
obey the probability distribution
$\ee^{-\frac{1}{4} \epsilon \sum_t \, \eta_k(t)^2}$.
Let us rescale them as
$\tilde{\eta}_k = \sqrt{\epsilon} \eta_k$ so that they obey the 
probability distribution
$\ee^{-\frac{1}{4} \sum_t \, \tilde{\eta}_k(t)^2}$ and hence, in particular,
\begin{alignat}{3}
\Big\langle \tilde{\eta}_k(t_1) \, \tilde{\eta}_l (t_2)  \Big\rangle_{\eta}
 = 2 \, \delta_{kl} \, \delta_{t_1  , t_2} \ .
\label{2pt-corr-eta-discretized}
\end{alignat}
With this normalization, the discretized Langevin equation 
(\ref{eq:Langevin-discretized}) becomes
\begin{alignat}{3}
x_k^{(\eta)} (t+\epsilon) 
= x_k^{(\eta)} (t) 
- \epsilon \frac{\del S}{\del x_k} 
+ \sqrt{\epsilon} \, \tilde{\eta}_k(t)  \ .
\label{eq:Langevin-discretized2}
\end{alignat}
Below we omit the tilde on $\eta$ to simplify the notation.

With this discretized version, we can
derive the FP equation (\ref{FPeq})
in a more elementary manner
than in the continuum \cite{Damgaard:1987rr}.
Let us consider a test function $f(x)$ and its expectation value
\begin{alignat}{3}
\Big\langle f \Big(x^{(\eta)}(t) \Big) \Big\rangle_{\eta}
=  \int dx f(x) P(x;t) 
\label{eq:f-P}
\end{alignat}
at a fictitious time $t$.
The $t$-evolution of this quantity is given by
\begin{alignat}{3}
& \Big\langle 
 f\Big(x^{(\eta)}(t+\epsilon)\Big) \Big\rangle_{\eta}
- \Big\langle f \Big(x^{(\eta)}(t)\Big) \Big\rangle_{\eta}  \nonumber \\
=&  \left\langle 
\frac{\del f}{\del x_k} \left( - \epsilon \frac{\del S}{\del x_k}\right)
+ \frac{1}{2} \frac{\del^2 f}{\del x_k \del x_l} (\sqrt{\epsilon} )^2 
\eta_k (t) \eta_l (t)
\right\rangle_{\eta} + O(\epsilon^2)  \nonumber \\
=& \epsilon \int dx 
\left(- \frac{\del f}{\del x_k}\frac{\del S}{\del x_k} 
+ \frac{\del^2 f}{\del x_k^2}
\right) P(x;t) + O(\epsilon^2) \nonumber \\
=& \epsilon \int dx  f(x)
\frac{\del}{\del x_k} 
\left(\frac{\del S}{\del x_k} + \frac{\del}{\del x_k} \right) P 
 + O(\epsilon^2) \ .
\label{eq:f-P-evolve}
\end{alignat}
Here we have used
\begin{alignat}{3}
  \left\langle 
\frac{1}{2} \frac{\del^2 f}{\del x_k \del x_l} 
(\sqrt{\epsilon})^2  \eta_k(t) \eta_l(t)
\right\rangle_{\eta} 
= \frac{1}{2} \epsilon  \left\langle 
 \frac{\del^2 f}{\del x_k \del x_l} \right\rangle_{\eta} 
\Big\langle   \eta_k(t)  \eta_l(t) \Big\rangle_{\eta} 
= \epsilon  \left\langle 
 \frac{\del^2 f}{\del x_k^2}\right\rangle_{\eta}  \ ,
\label{eq:a-trick}
\end{alignat}
which follows from the fact that
the function $\frac{\del^2 f}{\del x_k \del x_l}$ 
is evaluated at $x=x^{(\eta)}(t)$,
which depends only on $\eta(0)$, $\eta(\epsilon)$,
$\cdots$, $\eta(t-\epsilon)$, but not on $\eta(t)$.
Using (\ref{eq:f-P}), the same quantity (\ref{eq:f-P-evolve})
should be written as
\begin{alignat}{3}
\Big\langle f\Big(x^{(\eta)}(t+\epsilon)\Big) \Big\rangle_{\eta}
- \Big\langle f\Big(x^{(\eta)}(t)\Big) \Big\rangle_{\eta}
&=  \int dx f(x) \Big( P(x;t+\epsilon)  - P(x;t) \Big) \ .
\label{eq:f-P-evolve-P}
\end{alignat}
Since (\ref{eq:f-P-evolve}) and (\ref{eq:f-P-evolve-P}) should be
equal for an arbitrary $f(x)$, one obtains
\begin{alignat}{3}
P(x;t+\epsilon)  - P(x;t)
= \epsilon \frac{\del}{\del x_k} 
\left(\frac{\del S}{\del x_k} + \frac{\del}{\del x_k} \right) P 
 + O(\epsilon^2) \ .
\label{FPeq-discretized}
\end{alignat}
Thus, in the $\epsilon \rightarrow 0$ limit, one obtains (\ref{FPeq}).

\subsection{the case of complex action}
\label{sec:complex-action-case}
Let us apply the same method to the case in which
the action $S$ is a complex-valued function 
of the real variables $x_k$ ($k=1,\cdots ,n$).
In that case, however, 
the first term on the right-hand side of the Langevin equation 
(\ref{eq:Langevin}) becomes complex,
which means that $x_k^{(\eta)} (t)$ becomes complex
even if one starts from a real configuration
$x_k^{(\eta)} (0) \in \bbR$.
Let us therefore complexify the variables\footnote{In this 
respect, there is a closely related approach based on
the so-called Lefschetz 
thimble \cite{Witten:2010cx,Cristoforetti:2012su},
which has attracted much attention recently.
See refs.\cite{Cristoforetti:2013wha,%
Fujii:2013sra,Mukherjee:2014hsa,DiRenzo:2015foa,Fukushima:2015qza} 
and references therein.}
as $x_k  \mapsto z_k = x_k + i y_k$, and solve the
complex Langevin equation 
\begin{alignat}{3}
\dot{z}_k^{(\eta)} (t) = - \frac{\del S}{\del z_k} + \eta_k(t) \ ,
\label{eq:complex-Langevin}
\end{alignat}
where the action $S$ is now considered as a function 
of the complex variables $z_k$ ($k=1,\cdots ,n$)
by analytic continuation.
It is important for the method that the action $S(z)$
thus obtained is 
a holomorphic function of $z_k$.
The probabilistic variables $\eta_k(t)$ in 
(\ref{eq:complex-Langevin}) are, in general, complex
\begin{alignat}{3}
\eta_k(t)=\eta^{({\rm R})}_k(t)+ i \eta^{({\rm I})}_k(t) \ ,
\label{eq:complex-noise}
\end{alignat}
and obey the probability distribution
$\ee^{-\frac{1}{4} \int dt \, 
\{ \frac{1}{N_{\rm R}}\eta_k^{({\rm R})}(t)^2
+\frac{1}{N_{\rm I}}\eta_k^{({\rm I})}(t)^2 \} } $.
The probability distribution corresponding to (\ref{def-P}) 
is defined as
\begin{alignat}{3}
P(x,y;t) = \Bigl\langle \prod_k \delta \Big(x_k - x_k^{(\eta)} (t) \Big)
\, \delta \Big(y_k - y_k^{(\eta)} (t) \Big)
\Bigr \rangle_\eta \ ,
\label{def-P-xy}
\end{alignat}
where the expectation value $\langle \ \cdots \ \rangle_{\eta}$
is defined by
\begin{alignat}{3}
\langle \ \cdots \ \rangle_{\eta}
= \frac{\int {\cal D}\eta \cdots 
\ee^{
-\frac{1}{4} \int dt \, 
\{ \frac{1}{N_{\rm R}}\eta_k^{({\rm R})}(t)^2
+\frac{1}{N_{\rm I}}\eta_k^{({\rm I})}(t)^2 \}
}
}
{\int {\cal D}\eta  \, \ee^{-\frac{1}{4} \int dt \, 
\{ \frac{1}{N_{\rm R}}\eta_k^{({\rm R})}(t)^2
+\frac{1}{N_{\rm I}}\eta_k^{({\rm I})}(t)^2 \}
}} 
\ .
\label{def-EV-eta-complex}
\end{alignat}
With this notation, we have, for instance,
\begin{alignat}{3}
\Big\langle \eta^{({\rm R})}_k(t_1) \, \eta^{({\rm R})}_l (t_2)  
\Big\rangle_{\eta}
&= 2 N_{\rm R} \, \delta_{kl} \, \delta(t_1 - t_2) \ , \nonumber \\
\Big\langle \eta^{({\rm I})}_k(t_1) \, \eta^{({\rm I})}_l (t_2)  
\Big\rangle_{\eta}
&= 2 N_{\rm I} \, \delta_{kl} \, \delta(t_1 - t_2) \ , \nonumber \\
\Big\langle \eta_k^{({\rm R})}(t_1) \, \eta^{({\rm I})}_l (t_2)  
\Big\rangle_{\eta}
&= 0 \ .
\label{2pt-corr-eta-complex}
\end{alignat}
In what follows we assume
\begin{alignat}{3}
N_{\rm R} -N_{\rm I} = 1  
\label{NR-NI}
\end{alignat}
for a reason that becomes clear later.
For practical purposes, one should actually use
$N_{\rm R}=1$, $N_{\rm I} = 0$,
corresponding to real $\eta_k (t)$ with the distribution 
(\ref{def-EV-eta}),
to reduce the excursion
in the imaginary directions \cite{Aarts:2009uq,Aarts:2011ax}, which
spoils the validity of the method as we review below.

Repeating the analysis given 
in section \ref{sec:discretized-Langevin},
one can easily show that
$P(x,y;t)$ satisfies the FP-like equation
\begin{alignat}{3}
\frac{\del P}{\del t}
= \frac{\del}{\del x_k} 
\left\{ {\rm Re} \left( \frac{\del S}{\del z_k}\right) 
+ N_{\rm R} \frac{\del}{\del x_k} \right\} P 
+ \frac{\del}{\del y_k} 
\left\{ {\rm Im} \left(\frac{\del S}{\del z_k}\right) 
 +  N_{\rm I} \frac{\del}{\del y_k}
 \right\}  P \ .
\label{FP-like-eq}
\end{alignat}
In fact, for observables ${\cal O}(x)$ 
that admit holomorphic extension to ${\cal O}(x+iy)$,
one can show under certain conditions that
there exists a complex function $\rho(x;t)$, which satisfies
\begin{alignat}{3}
\int dx dy \, {\cal O}(x+iy) P(x,y;t)
= \int dx \, {\cal O}(x) \rho(x;t) \ ,
\label{P-rho-rel}
\end{alignat}
and obeys the differential equation
\begin{alignat}{3}
\frac{\del \rho}{\del t}
&= \frac{\del}{\del x_k} 
\left( \frac{\del S}{\del x_k} + 
\frac{\del}{\del x_k} \right) \rho \ ,
\label{FPeq-complex}
\end{alignat}
which is formally the same as the FP equation (\ref{FPeq}).
%
Although there is an important difference that
the action $S$ and $\rho$ are now complex,
the FP equation (\ref{FPeq-complex})
still has a time-independent solution
\begin{alignat}{3}
\rho_{\rm time-indep}(x) = \frac{1}{Z} \, \ee^{-S(x)} \ .
\label{time-indep-sol-complex}
\end{alignat}
The convergence
to this solution in the $t\rightarrow \infty$ limit
requires that
all the eigenvalues of 
the operator
acting on $\rho$ on the right-hand side of (\ref{FPeq-complex})
should have strictly negative real part except for the zero eigenvalue
corresponding to (\ref{time-indep-sol-complex}).
While this is not guaranteed in general unlike in the real action case,
one can argue that the convergence to 
(\ref{time-indep-sol-complex}) should occur if
the relation (\ref{P-rho-rel}) holds and 
the solution to the FP-like equation (\ref{FP-like-eq})
converges to some function uniquely.
Suppose the operator acting on $\rho$ has an eigenvalue with
a positive real part. Then the overall magnitude of $\rho$ increases 
exponentially with $t$, and (\ref{P-rho-rel}) cannot be satisfied.
Also, suppose the operator acting on $\rho$ has an eigenvalue with
a vanishing real part other than the zero eigenvalue
corresponding to (\ref{time-indep-sol-complex}).
Then the asymptotic behavior of $\rho$ 
depends on
the initial condition,
and (\ref{P-rho-rel}) cannot be satisfied with $P$ having
a unique asymptotic behavior.
To the best of our knowledge, this argument has been given 
for the first time in ref.~\cite{Nishimura:2015pba} with explicit
examples.

Thus, provided that the relation (\ref{P-rho-rel}) holds and 
the solution to the FP-like equation (\ref{FP-like-eq})
converges to some function uniquely,
we can calculate the VEV
with respect to the partition function (\ref{eq:part-fn}) as
\begin{alignat}{3}
\langle {\cal O} \rangle 
&= \int 
dx
\, {\cal O}(x) \, \rho_{\rm time-indep}(x) \nonumber\\
&= \lim_{t \rightarrow \infty} \int 
dx
\, {\cal O}(x) \, \rho(x;t) 
\nonumber\\
&= \lim_{t \rightarrow \infty} \int 
dx \, dy
\, 
{\cal O}(x+iy) \, P(x,y;t) \nonumber\\
&= \lim_{t \rightarrow \infty} 
\Big\langle  {\cal O}\Big(x^{(\eta)} (t)+i y^{(\eta)} (t)\Big)
\Big\rangle_{\eta} \nonumber \\
&= \lim_{T \rightarrow \infty} \frac{1}{T}
\int_{t_0}^{t_0+T} dt \, {\cal O}\Big(x^{(\eta)} (t)+i y^{(\eta)} (t)\Big) \ .
\label{O-time-av-complex}
\end{alignat}


In what follows, we
review the derivation\footnote{For an earlier work on this issue,
see ref.~\cite{Ambjorn:1985iw}.} of
the key relation (\ref{P-rho-rel}) given in 
refs.~\cite{Aarts:2009uq,Aarts:2011ax}.
At $t=0$, we can choose 
\begin{alignat}{3}
P(x,y;0)=\rho(x;0) \, \delta(y)  \ ,
\label{P-rho-initial}
\end{alignat}
where $\rho(x;0) \ge 0$ so that
(\ref{P-rho-rel}) holds trivially.
In order to prove the relation (\ref{P-rho-rel}) at arbitrary $t>0$,
we are going to show that
each side of (\ref{P-rho-rel}) can be rewritten as
\begin{alignat}{3}
\int dx dy \, {\cal O}(x+iy) \, P(x,y;t)
&= \int dx dy \, {\cal O}(x+iy;t) \, P(x,y;0) \ ,
\label{OP}
\\
\int dx \, {\cal O}(x) \, \rho(x;t)
&= \int dx \, {\cal O}(x;t) \, \rho(x;0)  \ .
\label{Orho}
\end{alignat}
In eq.~(\ref{OP}),
we have introduced the time-dependent observables
${\cal O}(z;t)$ defined by solving
\begin{alignat}{3}
\frac{\del}{\del t} {\cal O}(z;t) &= \tilde{L} \,  {\cal O}(z;t) \ , 
\label{Ot-def} \\
\tilde{L} 
&= \left( \frac{\del}{\del z_k} - \frac{\del S}{\del z_k} 
\right)
\frac{\del }{\del z_k}
\label{L-tilde}
\end{alignat}
with the initial condition
\begin{alignat}{3}
{\cal O}(z;0) &= {\cal O}(z) \ .
\label{Ot-initial}
\end{alignat}
Let us recall that we are considering 
holomorphic observables ${\cal O}(z)$.
One can actually show that 
the time-evolved observables ${\cal O}(z;t)$ 
remain to be holomorphic 
when $S(z)$ is a holomorphic function \cite{Aarts:2009uq}.
The observables ${\cal O}(x;t)$ that 
appear in (\ref{Orho})
are obtained by setting $y=0$ in
${\cal O}(x+iy;t)$, and they 
satisfy the differential equation
\begin{alignat}{3}
\frac{\del}{\del t} {\cal O}(x;t) &= L_0  {\cal O}(x;t) \ , 
\label{Oxt} \\
L_0 &= \left(
\frac{\del}{\del x_k} - \frac{\del S}{\del x_k} 
\right)
\frac{\del}{\del x_k} \ .
\label{L0-expression}
\end{alignat}
Since the right-hand sides of (\ref{OP}) and (\ref{Orho})
are equal to each other due to (\ref{P-rho-initial}), 
eqs.~(\ref{OP}) and (\ref{Orho}) imply
the desired relation (\ref{P-rho-rel}).

In order to show (\ref{OP}), we introduce the function
\begin{alignat}{3}
F(t,\tau) = \int dx dy \, {\cal O}(x+iy;\tau) \, P(x,y;t-\tau) \ ,
\label{def-F}
\end{alignat}
which interpolates each side of (\ref{OP}) with $0 \le \tau \le t$.
Taking the derivative with respect to $\tau$, we get
\begin{alignat}{3}
\frac{\del}{\del \tau} F(t,\tau) 
= \int dx dy \, \tilde{L} {\cal O}(x+iy;\tau) P(x,y;t-\tau) 
- \int dx dy \,  {\cal O}(x+iy;\tau) L^\top P(x,y;t-\tau) \ ,
\label{del-F}
\end{alignat}
where $L^\top$ denotes the operator acting on $P$
on the right-hand side of (\ref{FP-like-eq}).
The operator $L$ is then defined as an operator
satisfying $\langle Lf,g \rangle=\langle f,L^{\top} g \rangle$,
where $\langle f, g \rangle  \equiv \int f(x,y)g(x,y)dxdy$,
assuming that $f$ and $g$ are 
functions that allow integration by parts.
The explicit form of the operator $L$ can be obtained
as
\begin{alignat}{3}
L &= \left\{
 - {\rm Re} \left(\frac{\del S}{\del z_k}\right)
+ N_{\rm R} \frac{\del}{\del x_k}
\right\}
\frac{\del}{\del x_k}
+ \left\{
-   {\rm Im} \left(\frac{\del S}{\del z_k}\right) 
+ N_{\rm I} \frac{\del}{\del y_k}
\right\}
\frac{\del}{\del y_k} \ .
\label{L-expression}
\end{alignat}
An important observation here is that when
$L$ acts on a holomorphic function $f(z)$ of $z_k$, 
it can be replaced by $\tilde{L}$ since
\begin{alignat}{3}
L f(z) &= \left\{
- {\rm Re} \left(\frac{\del S}{\del z_k}\right)
+ N_{\rm R} \frac{\del}{\del z_k} 
\right\}
\frac{\del f}{\del z_k}
+ \left\{
 -  {\rm Im} \left(\frac{\del S}{\del z_k}\right) 
+ i N_{\rm I} \frac{\del}{\del z_k}
\right\}
\left( i \frac{\del f}{\del z_k} \right)  \nonumber \\
&= \left\{
-  \frac{\del S}{\del z_k}
+ ( N_{\rm R} - N_{\rm I})  \frac{\del}{\del z_k} 
\right\}
\frac{\del f}{\del z_k}
\nonumber \\
 &= \tilde{L} f(z) \ ,
\label{Lf}
\end{alignat}
where we have used (\ref{NR-NI}).
This implies that $\tilde{L}$ in the first term of 
(\ref{del-F}) can be replaced by $L$,
and hence (\ref{del-F}) vanishes
if one can perform integration by parts.
In that case, $F(t,\tau)$ is independent of $\tau$,
and (\ref{OP}) follows.

A similar argument can be used to show (\ref{Orho}).
We define
\begin{alignat}{3}
G(t,\tau) = \int dx \, {\cal O}(x;\tau) \, \rho(x;t-\tau) \ ,
\label{def-G}
\end{alignat}
which interpolates each side of (\ref{Orho}) for $0 \le \tau \le t$.
Taking the derivative with respect to $\tau$, we get
\begin{alignat}{3}
\frac{\del}{\del \tau} G(t,\tau) 
= \int dx  \, L_0 {\cal O}(x;\tau) \rho(x;t-\tau) 
- \int dx  \,  {\cal O}(x;\tau) L_0^\top \rho(x;t-\tau) \ ,
\label{del-G}
\end{alignat}
where 
we have used (\ref{Oxt}) and (\ref{FPeq-complex}).
Here the integration on the right-hand side involves the
real directions $x_k$ only, so we can 
perform integration by parts without any problem
due to the effects of the action, which make 
$\rho(x;t)$ well localized.
Thus (\ref{del-G}) vanishes,
and (\ref{Orho}) follows.

On the other hand, the integration by parts that one needs to use
to show that (\ref{del-F}) vanishes involves the 
imaginary directions $y_k$.
It can therefore be justified only if
the probability distribution
$P(x,y;t)$ has a sharp fall-off 
in the imaginary directions \cite{Aarts:2009uq,Aarts:2011ax}.

Recently it has been pointed out
that the integration by parts can be invalidated also 
when the drift term includes a singularity \cite{Nishimura:2015pba}.
This issue is relevant, in particular, to complex action systems involving
fermions such as finite density QCD since the fermion determinant
gives rise to a singular drift term.
The CLM still works if
the probability distribution
$P(x,y;t)$ is suppressed strongly enough near the singularity.


\section{``Gauge cooling'' in 0-dimensional systems}
 \label{sec:cooling}

At the end of the previous section, we discussed two possible
problems, which can make the CLM give wrong results.
The gauge cooling was originally proposed to cure the first problem
\cite{Seiler:2012wz} in gauge theories. 
In ref.~\cite{NNS} we propose 
that it can be applied also to cure the second problem,
and demonstrate that it does in the Random Matrix Theory
for finite density QCD.
In this section we consider the ``gauge cooling''
in 0-dimensional systems such as the Random Matrix Theory
and provide an explicit justification.
Apart from pedagogical purposes, we consider that
it
is useful, in particular, in matrix models relevant to 
superstring theory \cite{Kim:2011cr,Anagnostopoulos:2013xga}.
Generalization to the lattice gauge theory is
straightforward and it is given in 
sections \ref{sec:lattice} and \ref{sec:cooling-lgt}.

Let us consider a system of $N$ real variables
with a symmetry under
\begin{alignat}{3}
x_j ' = g_{jk} x_k \ ,
\label{symmetry}
\end{alignat}
where $g$ is a representation matrix of a Lie group.
An infinitesimal transformation is denoted as
\begin{alignat}{3}
\delta x_j = i \lambda_{jk} x_k \ .
\label{symmetry-infinitesimal}
\end{alignat}
Here $\lambda$ is an element of the Lie algebra,
which can be expanded as
\begin{alignat}{3}
\lambda_{jk} = \sum_a \lambda_a (t_a)_{jk} 
\label{lambda-expand}
\end{alignat}
in terms of 
the generators $t_a$ of the Lie group under consideration
with real coefficients $\lambda_a \in \bbR$ .
Upon complexifying the variables 
$x_k  \mapsto z_k = x_k + i y_k$,
the symmetry of the action and the observables 
naturally enhances from (\ref{symmetry}) to
\begin{alignat}{3}
z_j ' = g_{jk} z_k \ ,
\label{symmetry-complexified}
\end{alignat}
where $g$ is an element of the Lie group that can be obtained
by complexifying the original Lie group.
In particular, an infinitesimal transformation of the complexified symmetry 
is given by
\begin{alignat}{3}
\delta z_j = i \lambda_{jk} z_k \ .
\label{symmetry-infinitesimal-complexified}
\end{alignat}
Here $\lambda$ is an element of the Lie algebra
for the complexified Lie group, which can be expanded
as (\ref{lambda-expand}) but now with complex coefficients
$\lambda_a \in \bbC$.

As a simple example, let us consider an O($N$) vector model
\begin{alignat}{3}
S(x) = 
\sigma \sum_{k=1}^N (x_k)^2 
+ \kappa \Big\{\sum_{k=1}^N (x_k)^2 \Big\}^2 
\label{On-model}
\end{alignat}
with $\sigma \in \bbC$,
which is invariant under (\ref{symmetry}) with $g\in {\rm O}(N)$.
An infinitesimal transformation is given by
(\ref{symmetry-infinitesimal}),
where $\lambda_{jk}$ is 
a purely imaginary antisymmetric $N\times N$ matrix.
Upon complexification $x_k  \mapsto z_k = x_k + i y_k$,
the action becomes
\begin{alignat}{3}
S(z) = 
\sigma \sum_{k=1}^N (z_k)^2 
+ \kappa \Big\{\sum_{k=1}^N (z_k)^2 \Big\}^2 \ ,
\label{On-model-complexified}
\end{alignat}
which is invariant under (\ref{symmetry-complexified})
with $g\in {\rm O}(N,\bbC)$,
namely with $g$ being an $N \times N$ complex matrix
satisfying $g ^{\top} g = g g ^{\top} = {\bf 1}$.
(The symbol $g ^{\top}$ here represents the transpose of the matrix $g$.)
An infinitesimal transformation is given by
(\ref{symmetry-infinitesimal-complexified}),
where $\lambda_{jk}$ is 
a complex antisymmetric $N\times N$ matrix.

The discretized version of the
complex Langevin equation (\ref{eq:complex-Langevin})
can be written as
\begin{alignat}{3}
z_k^{(\eta)} (t+\epsilon) 
= z_k^{(\eta)} (t) 
- \epsilon \frac{\del S}{\del z_k} 
+ \sqrt{\epsilon} \, \eta_k(t)  \ ,
\label{eq:Langevin-discretized2-complexified}
\end{alignat}
analogously to (\ref{eq:Langevin-discretized2}).
The probabilistic variables 
\begin{alignat}{3}
\eta_k(t)=\eta^{({\rm R})}_k(t)
+ i \eta^{({\rm I})}_k(t) 
\label{eq:complex-noise-discretized}
\end{alignat}
obey the probability distribution
$\ee^{-\frac{1}{4} \sum_t 
\{ \frac{1}{N_{\rm R}}\eta_k^{({\rm R})}(t)^2
+\frac{1}{N_{\rm I}}\eta_k^{({\rm I})}(t)^2 \} } $.
The gauge cooling \cite{Seiler:2012wz}
is a procedure of making a complexified
symmetry transformation (\ref{symmetry-complexified})
between the Langevin steps.
Thus it amounts to modifying the complex Langevin equation 
(\ref{eq:Langevin-discretized2-complexified}) into
\begin{alignat}{3}
\tilde{z}_k^{(\eta)} (t) &= 
g_{kl}
\, z_l^{(\eta)} (t) \ ,  \nonumber \\
z_k^{(\eta)} (t+\epsilon) 
&=   \tilde{z}_k^{(\eta)} (t) 
- \epsilon \frac{\del S(\tilde{z})}{\del \tilde{z}_k} 
+ \sqrt{\epsilon} \, \eta_k(t)  \ ,
\label{eq:Langevin-discretized2-complexified-cooled}
\end{alignat}
where $g$ is an element of the complexified Lie group
chosen appropriately as a function of the configuration before cooling.
The basic idea is to determine $g$
in such a way that the modified Langevin process 
(\ref{eq:Langevin-discretized2-complexified-cooled})
does not suffer from the problem of the original 
Langevin process (\ref{eq:Langevin-discretized2-complexified}).
Clearly this idea will have a chance to work only if
the degrees of freedom in the symmetry transformation
have at least 
the same order of magnitude as those of the dynamical system itself.
Gauge theories are one such example, but 0-dimensional
models such as vector models and matrix models
would be equally good as demonstrated 
explicitly in the Random Matrix Theory \cite{NNS}.

For instance, if the excursions in the imaginary directions
are problematic in studying the model (\ref{On-model}) by the CLM, 
one can introduce the norm\footnote{This is analogous to 
the so-called ``unitarity norm'' \cite{Seiler:2012wz} 
proposed in the complex Langevin simulation 
of lattice gauge theory.}
\begin{alignat}{3}
{\cal N} = 
\sum_{k=1}^N  ( y_k )^2
= -
\frac{1}{4} \sum_{k=1}^N (z_k - z_k^*)^2 
\ , 
\label{hermiticity-norm}
\end{alignat}
which measures the distance from the real region,
and determine the transformation $g$
in (\ref{eq:Langevin-discretized2-complexified-cooled})
in such a way that
the norm ${\cal N}$ for 
$z^{(\eta)} (t)$
is reduced by the transformation.
In ref.~\cite{NNS} we propose to combine this norm
with another norm to cure also the problem caused 
by a singular drift term.
Typically the norm one tries to reduce is invariant under
transformations in the original Lie group but not under
transformations in the complexified Lie group
as in the case of (\ref{hermiticity-norm}).
The main issue we address below is
whether the modification of the Langevin process
by the ``gauge cooling'' spoils the equivalence
to the path integral reviewed in section \ref{sec:complex-action-case}.

Note that the gauge cooling is a completely deterministic procedure.
In particular, the transformation $g$ in
(\ref{eq:Langevin-discretized2-complexified-cooled})
is determined only by the configuration
$z^{(\eta)}(t)$ before cooling.
Therefore, for our purpose, 
it is convenient to rewrite 
(\ref{eq:Langevin-discretized2-complexified-cooled})
into the form
\begin{alignat}{3}
z_k^{(\eta)} (t+\epsilon) 
&= g_{kl}
\left( z_l^{(\eta)} (t) 
- \epsilon \frac{\del S}{\del z_l} \right)
+ \sqrt{\epsilon} \eta_k(t)  \ ,
\label{eq:Langevin-discretized2-complexified-cooled-tilde}
\end{alignat}
which involves $z^{(\eta)}(t)$ only.\footnote{In practice,
one usually measures observables using the configuration
$\tilde{z}^{(\eta)}(t)$ after cooling instead of $z^{(\eta)}(t)$.
This does not cause any problem 
since $z^{(\eta)}(t)$ and $\tilde{z}^{(\eta)}(t)$ are
related to each other 
by the complexified symmetry transformation (\ref{symmetry-complexified}),
under which the observables are invariant.
}
Let us assume that 
$g$ is given by
\begin{alignat}{3}
g
& = 
\exp \Big\{  i \epsilon 
\lambda(x^{(\eta)} (t),y^{(\eta)} (t)) 
 \Big\} \ , 
\label{lambda-def}
\end{alignat}
where 
$\lambda(x,y)$ is an element of the Lie algebra of 
${\rm O}(N,\bbC)$.
For instance, one may use
\begin{alignat}{3}
\lambda_{kl}(x,y) &=
 \alpha(x,y)  (x_k y_l - y_k x_l) 
= \frac{1}{2} \,  i \alpha(x,y)
(z_k z_l^{*} - z_k^{*} z_l) \ , 
\label{eq:gradient}
\end{alignat}
which can be obtained by calculating the gradient of 
the norm (\ref{hermiticity-norm}) with respect to
the ${\rm O}(N,\bbC)$ transformation.
The real positive function $\alpha(x,y)$ 
can be chosen to optimize the reduction of the norm.
(In practice, one can simply determine $g$ 
by using the steepest descent method
with respect to the norm (\ref{hermiticity-norm}) \cite{Seiler:2012wz}.)
Note that $\lambda(x,y)$ is not a holomorphic function of $z_k$
in general as in (\ref{eq:gradient}).

Using (\ref{lambda-def})
in eq.~(\ref{eq:Langevin-discretized2-complexified-cooled-tilde})
and taking the $\epsilon \rightarrow 0$ limit,
we obtain the continuum complex Langevin equation 
for $z^{(\eta)}(t)$ as
\begin{alignat}{3}
\dot{z}_k^{(\eta)} (t) = 
- \frac{\del S}{\del z_k} + \eta_k(t) 
+ i \lambda_{kl}\Big(x^{(\eta)} (t),y^{(\eta)} (t)\Big)
 \, z_l^{(\eta)} (t) \ ,
\label{eq:complex-Langevin-cooling}
\end{alignat}
where the effect of the gauge cooling is represented by the
last term on the right-hand side. 
Then we can easily find that the FP-like equation (\ref{FP-like-eq}) 
that $P(x,y;t)$ satisfies is modified by the gauge cooling as
\begin{alignat}{3}
\frac{\del P}{\del t}
=& \frac{\del}{\del x_k} 
\left\{ {\rm Re} \left( \frac{\del S}{\del z_k}
-i  \lambda_{kl}(x,y) z_l \right) 
+ N_{\rm R} \frac{\del}{\del x_k} \right\} P  \nonumber
\\
&+ \frac{\del}{\del y_k} 
\left\{ {\rm Im} \left(\frac{\del S}{\del z_k}
-i  \lambda_{kl}(x,y) z_l \right) 
+ N_{\rm I} \frac{\del}{\del y_k} 
  \right\}  P \ .
\label{FP-like-eq-cooling}
\end{alignat}
This modifies the differential operator $L$ 
in eq.~(\ref{L-expression}) into
\begin{alignat}{3}
L =& \left\{
 - {\rm Re} \left(\frac{\del S}{\del z_k} 
-i  \lambda_{kl}(x,y) z_l 
\right)
+N_{\rm R} \frac{\del}{\del x_k}
\right\}
\frac{\del}{\del x_k}
\nonumber \\
& + \left\{
-   {\rm Im} \left(\frac{\del S}{\del z_k}
-i  \lambda_{kl}(x,y) z_l 
\right) 
+ N_{\rm I} \frac{\del}{\del y_k}
\right\}
\frac{\del}{\del y_k} \ .
\label{L-expression-mod}
\end{alignat}
Acting this modified $L$
on a holomorphic function $f(z)$,
we obtain
\begin{alignat}{3}
L f(z) 
&=
\left(
 - \frac{\del S}{\del z_k} 
+i  \lambda_{kl}(x,y) z_l 
 +
(N_{\rm R}-N_{\rm I})
\frac{\del}{\del z_k}
\right)
\frac{\del}{\del z_k} f(z)
\nonumber \\
&= \tilde{L} f(z)
+i  \lambda_{kl}(x,y) z_l \frac{\del}{\del z_k}  f(z) \ ,
\label{Lf-modified}
\end{alignat}
where $\tilde{L}$ is defined by (\ref{L-tilde}).
The extra term compared with (\ref{Lf})
represents the change of $f(z)$
under an infinitesimal O($N,\bbC$) transformation.
Note that 
the time-evolved observables ${\cal O}(z;t)$ 
defined by (\ref{Ot-def})
remain to be invariant under the O($N,\bbC$) transformation
as long as the action $S(z)$
and the original observables ${\cal O}(z)$ are invariant.
Therefore, the $\tilde{L}$ in the first term of 
(\ref{del-F}) can be replaced by the modified $L$.
Hence (\ref{del-F}) vanishes
if one can perform integration by parts
for the modified $L$ and $P$.
In that case, the crucial identity (\ref{P-rho-rel}) 
holds for the modified $P$ with the same $\rho$.
Thus we have shown explicitly that the ``gauge cooling''
provides a possibility to improve 
the property of the probability distribution $P(x,y;t)$
so that (\ref{P-rho-rel}) holds, without affecting
the FP equation (\ref{FPeq-complex}) for $\rho(x;t)$.


\section{Application of the 
CLM
to lattice gauge theory}
\label{sec:lattice}

In this section we discuss
the application of the CLM
to lattice gauge theory, 
which is defined by the partition function
\begin{alignat}{3}
Z = \int dU \, \ee^{-S(U)} = \int \prod_{n \mu} dU_{n\mu} 
\, \ee^{-S(U)}  \ ,
  \label{eq:part-fn-lgt}
\end{alignat}
where the action $S$ is a complex-valued function 
of the configuration $U = \{ U_{n \mu} \}$
composed of link variables $U_{n \mu} \in {\rm SU}(3)$,
and the integration measure $dU_{n\mu}$ represents the 
Haar measure for the SU(3) group.
The only complication compared with the case discussed
in the previous sections
comes from the fact that the
dynamical variables take values on a group manifold.
The Langevin equation
in such a case with a real action is discussed intensively
in refs.~\cite{Alfaro:1982ef,Drummond:1982sk,%
Guha:1982uj,Halpern:1983jt,Batrouni:1985qr}.
Using this formulation, we can easily generalize
our discussions to the case of lattice gauge theory.

When the action $S$ is complex,
the drift term in the Langevin equation 
makes the link variables evolve into 
${\rm SL}(3,\bbC)$
matrices (i.e., $3\times 3$ general complex matrices 
with the determinant one)
even if one starts from a configuration of
${\rm SU}(3)$ matrices.
Let us therefore complexify the link variables as
${\cal U}_{n \mu} \in  {\rm SL}(3,\bbC)$, and solve the
complex Langevin equation 
\begin{alignat}{3}
\dot{{\cal U}}_{n \mu}^{(\eta)} (t) = 
i \sum_a 
\Big( - {\cal D}_{a n \mu} S({\cal U})
+ \eta_{a n \mu}(t) \Big)
\, t_a \, {\cal U}_{n \mu}^{(\eta)} (t) \ ,
\label{eq:complex-Langevin-lgt}
\end{alignat}
where the action $S({\cal U})$ is now considered as a holomorphic
function of the complexified configuration ${\cal U}_{n \mu}$,
and $t_a$ are the generators of the SU(3) group
normalized by $\tr (t_a t_b)=\delta_{ab}$.
The probabilistic variables $\eta_{a n \mu}(t)$ are 
defined similarly to (\ref{eq:complex-noise-discretized}).
The derivative operator ${\cal D}_{a n \mu}$ is defined 
as\footnote{The derivative operators defined in eqs.~(\ref{def-D-lgt-DR-DI})
and (\ref{def-Dbar-lgt}) may be regarded as analogues
of $\frac{\del}{\del z}= \frac{1}{2}(\frac{\del}{\del x}
- i \frac{\del}{\del y})$ and 
$\frac{\del}{\del \bar{z}}= \frac{1}{2}(\frac{\del}{\del x}
+ i \frac{\del}{\del y})$, respectively.}
\begin{alignat}{3}
{\cal D}_{a n \mu} &= 
\frac{1}{2}({\cal D}^{\rm (R)}_{a n \mu} - i {\cal D}^{\rm (I)}_{a n \mu})  \ ,
\label{def-D-lgt-DR-DI} \\
{\cal D}^{\rm (R)}_{a n \mu} f({\cal U})
&= \left. \frac{\del}{\del x} 
f(e^{i x t_a}  {\cal U}_{n \mu}) \right|_{x=0}  \ ,
\label{def-DR-lgt} \\
{\cal D}^{\rm (I)}_{a n \mu} f({\cal U})
&= \left. \frac{\del}{\del y} 
f(e^{- y t_a}  {\cal U}_{n \mu}) \right|_{y=0} \ .
\label{def-DI-lgt} 
\end{alignat}
Here $f({\cal U})$ are functions 
on the complexified group manifold, which
are not necessarily holomorphic, and $x$ and $y$ 
in eqs.~(\ref{def-DR-lgt}) and (\ref{def-DI-lgt})
are real parameters.
Note that for a holomorphic function $f({\cal U})$,
we have 
$\bar{\cal D}_{a n \mu} f({\cal U}) = 0 $, where
\begin{alignat}{3}
\bar{\cal D}_{a n \mu}   & =
\frac{1}{2}({\cal D}^{\rm (R)}_{a n \mu} + i {\cal D}^{\rm (I)}_{a n \mu}) 
\ ,
\label{def-Dbar-lgt}
\end{alignat}
and hence
\begin{alignat}{3}
 {\cal D}^{\rm (R)}_{a n \mu}  f({\cal U})  
 = {\cal D}_{a n \mu} f({\cal U}) \ ,
\quad \quad \quad 
 {\cal D}^{\rm (I)}_{a n \mu}  f({\cal U})  
 = i {\cal D}_{a n \mu} f({\cal U})  \ .
\label{holomorphic-func}
\end{alignat}

Then we define the probability distribution
\begin{alignat}{3}
P({\cal U};t) = \Bigl\langle \prod_{n \mu}
\delta \Big({\cal U}_{n \mu} , {\cal U}_{n \mu}^{(\eta)} (t) \Big)
\Bigr \rangle_\eta \ ,
\label{def-P-xy-lgt}
\end{alignat}
where the delta function is defined by
\begin{alignat}{3}
\int d  {\cal U}   \, 
f({\cal U}) \,
\delta \Big({\cal U}_{n \mu} ,  \widetilde{\cal U}_{n \mu}  \Big)
= f(\widetilde{\cal U})
\label{def-delta-lgt}
\end{alignat}
for any function $f({\cal U})$.
The integration measure that appears on the left-hand side
represents the Haar measure for the ${\rm SL}(3,\bbC)$ group
normalized appropriately.
One can show that the probability distribution
$P({\cal U};t)$
obeys the FP-like equation (See Appendix \ref{sec:derive-FP-like-eq-lgt}
for derivation.)
\begin{alignat}{3}
\frac{\del P}{\del t}
= & {\cal D}_{a n \mu}^{\rm (R)}
\left\{ 
{\rm Re} \Big( {\cal D}_{a n \mu} S ({\cal U}) \Big)
+ N_{\rm R} {\cal D}_{a n \mu}^{\rm (R)} \right\} P 
+ {\cal D}_{a n \mu}^{\rm (I)}
\left\{ 
{\rm Im} \Big( {\cal D}_{a n \mu} S ({\cal U}) \Big)
+ N_{\rm I} {\cal D}_{a n \mu}^{\rm (I)} \right\} P \ .
\label{FP-like-eq-lgt}
\end{alignat}
In fact, for observables ${\cal O}(U)$ 
that admit holomorphic extension to ${\cal O}({\cal U})$,
one can show under certain conditions that there exists
a complex function $\rho(U;t)$, which satisfies
\begin{alignat}{3}
\int d  {\cal U}   \, 
{\cal O}({\cal U}) \, P({\cal U};t)
= \int dU \, {\cal O}(U) \, \rho(U;t) \ ,
\label{P-rho-rel-lgt}
\end{alignat}
and obeys the FP equation
\begin{alignat}{3}
\frac{\del }{\del t}\rho(U;t)
&= D_{a n \mu}
\Big(  D_{a n \mu} S(U) + D_{a n \mu} \Big) \, 
\rho(U;t)  \ .
\label{FPeq-complex-lgt}
\end{alignat}
Here we have defined 
the derivative operator $D_{a n \mu}$,
which acts on a function 
$f(U)$ of the unitary gauge configuration as
\begin{alignat}{3}
D_{a n \mu} f(U)
= \left. \frac{\del}{\del x} 
f(e^{i x t_a}  U_{n \mu}   ) \right|_{x=0} \ .
\label{def-Dxi-lgt}
\end{alignat}
Note that the FP equation (\ref{FPeq-complex-lgt})
has a time-independent solution
\begin{alignat}{3}
\rho_{\rm time-indep}(U) = \frac{1}{Z} 
\exp \Big(-S(U) \Big) \ .
\label{time-indep-sol-complex-lgt}
\end{alignat}
As we argued in section \ref{sec:complex-action-case},
the convergence to 
(\ref{time-indep-sol-complex-lgt}) should occur if
the relation (\ref{P-rho-rel-lgt}) holds and 
the FP-like equation (\ref{FP-like-eq-lgt})
converges to some function uniquely.
In that case, we can calculate the VEV
with respect to the partition function (\ref{eq:part-fn-lgt}) as
\begin{alignat}{3}
\langle {\cal O} \rangle 
&= \int  dU \, {\cal O}(U) \rho_{\rm time-indep}(U) \nonumber \\
&= \lim_{t \rightarrow \infty} \int  dU \, {\cal O}(U) \rho(U;t) \nonumber\\
&= \lim_{t \rightarrow \infty} \int d{\cal U}  \, 
{\cal O}({\cal U}) P({\cal U};t) \nonumber \\
&= \lim_{t \rightarrow \infty} 
\Big\langle  {\cal O}\Big({\cal U}^{(\eta)} (t)\Big)
\Big\rangle_{\eta} \nonumber \\
&= \lim_{T \rightarrow \infty} \frac{1}{T}
\int_{t_0}^{t_0+T} dt \, {\cal O}\Big( {\cal U}^{(\eta)} (t)\Big) \ .
\label{O-time-av-complex-lgt}
\end{alignat}

Let us briefly discuss how one can 
derive the relation (\ref{P-rho-rel-lgt}).
At $t=0$, we choose 
\begin{alignat}{3}
P({\cal U},;0)=\int  dU \, \rho(U;0)
\prod_{n \mu}
\delta \Big({\cal U}_{n \mu} , U_{n \mu} \Big)
\label{P-rho-initial-lgt}
\end{alignat}
with $\rho(U;0) \ge 0$ so that
(\ref{P-rho-rel-lgt}) holds trivially.
In order to prove the relation (\ref{P-rho-rel-lgt}) at arbitrary $t>0$,
we are going to show that
each side of (\ref{P-rho-rel-lgt}) can be rewritten as
%
\begin{alignat}{3}
\int d  {\cal U}  \, 
{\cal O}({\cal U}) \, P({\cal U};t)
&= 
\int d  {\cal U}   \, 
{\cal O}({\cal U};t) \, P({\cal U};0) \ ,
\label{OP-lgt}
\\
\int dU \, {\cal O}(U) \, \rho(U;t) 
&= 
\int dU \, {\cal O}(U;t) \, \rho(U;0) \  .
\label{Orho-lgt}
\end{alignat}
In eq.~(\ref{OP-lgt}),
we have introduced the time-dependent observables
${\cal O}({\cal U};t)$ defined by solving
\begin{alignat}{3}
\frac{\del}{\del t} {\cal O}({\cal U};t) & = \tilde{L} \,
  {\cal O}({\cal U};t) \ ,  \label{Ot-def-lgt} \\
\tilde{L} &=
\Big( {\cal D}_{a n \mu} - {\cal D}_{a n \mu} S ({\cal U}) \Big)
{\cal D}_{a n \mu}  
\label{Lftilde-lgt}
\end{alignat}
with the initial condition
\begin{alignat}{3}
{\cal O}({\cal U};0) &= {\cal O}({\cal U}) \ .
\label{Ot-initial-lgt}
\end{alignat}
Let us recall that we are considering 
holomorphic observables ${\cal O}({\cal U})$.
One can actually show that 
the time-evolved observables ${\cal O}({\cal U};t)$ 
remain to be holomorphic 
when $S({\cal U})$ is a holomorphic function \cite{Aarts:2009uq}.
The observables ${\cal O}(U;t)$ that 
appears in (\ref{Orho-lgt})
are obtained by setting ${\cal U}=U$ in
${\cal O}({\cal U};t)$, and they 
satisfy the differential equation
\begin{alignat}{3}
\frac{\del}{\del t} {\cal O}(U;t) &= L_0  {\cal O}(U;t) \ , 
\label{Oxt-lgt} \\
L_0 &=
\Big(  D_{a n \mu} -D_{a n \mu} S(U) 
 \Big)
D_{a n \mu}  \ .
\label{L0-expression-lgt}
\end{alignat}
Since the right-hand sides of (\ref{OP-lgt}) and (\ref{Orho-lgt})
are equal to each other due to (\ref{P-rho-initial-lgt}), 
eqs.~(\ref{OP-lgt}) and (\ref{Orho-lgt}) imply
the desired relation (\ref{P-rho-rel-lgt}).
%

In order to show (\ref{OP-lgt}), we introduce the function
\begin{alignat}{3}
F(t,\tau) = 
\int d  {\cal U}   \, 
{\cal O}({\cal U};\tau) \, P({\cal U};t-\tau) \ ,
\label{def-F-lgt}
\end{alignat}
which interpolates each side of (\ref{OP-lgt}) with $0 \le \tau \le t$.
Taking the derivative with respect to $\tau$, we get
\begin{alignat}{3}
\frac{\del}{\del \tau} F(t,\tau) 
= 
\int d  {\cal U}  \, 
\tilde{L} {\cal O}({\cal U};\tau) \, P({\cal U};t-\tau) 
-
\int d  {\cal U}   \, 
{\cal O}({\cal U};\tau) L^\top  P({\cal U};t-\tau) \  ,
\label{del-F-lgt}
\end{alignat}
where $L^\top$ denotes the operator acting on $P$
on the right-hand side of (\ref{FP-like-eq-lgt}).
The operator $L$ is then defined as an operator
satisfying $\langle Lf,g \rangle=\langle f,L^{\top} g \rangle$,
where $\langle f, g \rangle  \equiv 
\int f({\cal U})
g({\cal U}) d{\cal U} $,
assuming that $f$ and $g$ are 
functions that allow integration by parts.
The explicit form of the operator $L$ can be obtained
as
\begin{alignat}{3}
L =& 
\left\{ 
- {\rm Re} \Big( {\cal D}_{a n \mu} S ({\cal U}) \Big)
+ N_{\rm R} {\cal D}_{a n \mu}^{\rm (R)}
\right\}
{\cal D}_{a n \mu}^{\rm (R)}  
+
\left\{ 
- {\rm Im} \Big( {\cal D}_{a n \mu} S ({\cal U}) \Big)
+ N_{\rm I} {\cal D}_{a n \mu}^{\rm (I)}
\right\}
{\cal D}_{a n \mu}^{\rm (I)}  \ . 
\label{L-expression-lgt}
\end{alignat}
An important observation here is that when
$L$ acts on a holomorphic function $f({\cal U})$,
it can be replaced by $\tilde{L}$ since
\begin{alignat}{3}
L f({\cal U}) =& 
\left\{ 
- {\rm Re} \Big( {\cal D}_{a n \mu} S ({\cal U}) \Big)
+ N_{\rm R} {\cal D}_{a n \mu}
\right\}
{\cal D}_{a n \mu} f({\cal U})
 \nonumber \\
& + \left\{ 
- {\rm Im} \Big( {\cal D}_{a n \mu} S ({\cal U}) \Big)
+ i N_{\rm I} {\cal D}_{a n \mu}
\right\}
i {\cal D}_{a n \mu} f({\cal U}) 
\nonumber \\
=& 
\left\{ - {\cal D}_{a n \mu} S ({\cal U})
+ (N_{\rm R}- N_{\rm I}) {\cal D}_{a n \mu}  \right\}
{\cal D}_{a n \mu}  f({\cal U})
\nonumber \\
=& \tilde{L} f({\cal U}) \ ,
\label{Lf-lgt}
\end{alignat}
where we have used (\ref{holomorphic-func}) and (\ref{NR-NI}).
This implies that $\tilde{L}$ in the first term of 
(\ref{del-F-lgt}) can be replaced by $L$,
and hence (\ref{del-F-lgt}) vanishes
if one can perform integration by parts.
In that case, $F(t,\tau)$ is independent of $\tau$,
and (\ref{OP-lgt}) follows.

A similar argument can be used to show (\ref{Orho-lgt}).
We define
\begin{alignat}{3}
G(t,\tau) = 
\int dU \, {\cal O}(U;\tau) \, \rho(U;t-\tau)
 \ ,
\label{def-G-lgt}
\end{alignat}
which interpolates each side of (\ref{Orho-lgt}) for $0 \le \tau \le t$.
Taking the derivative with respect to $\tau$, we get
\begin{alignat}{3}
\frac{\del}{\del \tau} G(t,\tau) 
= 
\int dU \, L_0 {\cal O}(U;\tau) \, \rho(U;t-\tau)
- \int dU \,  {\cal O}(U;\tau) L_0^\top \rho(U;t-\tau)  \ ,
\label{del-G-lgt}
\end{alignat}
where 
we have used (\ref{Oxt-lgt}) and (\ref{FPeq-complex-lgt}).
Here the integration on the right-hand side involves the
real directions only, which are compact in the present case,
so we can perform integration by parts without any problem
to show that (\ref{del-G-lgt}) vanishes.
Thus $G(t,\tau)$ is independent of $\tau$, and (\ref{Orho-lgt}) follows.

On the other hand, 
the integration by parts in (\ref{del-F-lgt}) is justified only if
the probability distribution
$P({\cal U};t)$ has a sharp fall-off in the 
noncompact imaginary directions.
The gauge cooling \cite{Seiler:2012wz} was originally
proposed to solve this problem.
As we mentioned in 
section \ref{sec:complex-action-case},
the integration by parts can be invalidated also 
when the drift term includes a singularity \cite{Nishimura:2015pba}.
This problem is anticipated to occur
when one applies the CLM to
finite density QCD at low temperature with light quarks.
The CLM still works if
the probability distribution
$P({\cal U};t)$ 
is suppressed strongly enough near the singularity.
We consider that the gauge cooling is useful also in
solving this problem as in the case of Random Matrix Theory \cite{NNS}.

\section{Gauge cooling in lattice gauge theory}
 \label{sec:cooling-lgt}

The lattice gauge theory is invariant 
under the ${\rm SU}(3)$ gauge transformation.
For instance, the plaquette action
\begin{alignat}{3}
S_{\rm plaquette}(U) = - \beta \sum_{n}  \sum_{\mu \neq \nu}
\tr (U_{n\mu} U_{n+\hat{\mu},\nu} 
U_{n+\hat{\nu},\mu}^{-1} U_{n\nu}^{-1} ) 
\label{plaquette-action}
\end{alignat}
is invariant under
\begin{alignat}{3}
U_{n \mu} ' = g_{n} U_{n \mu} g_{n+\hat{\mu}}^{-1} \ ,
\label{symmetry-lgt}
\end{alignat}
where $g_{n} \in {\rm SU}(3)$.
An infinitesimal transformation is denoted as
\begin{alignat}{3}
\delta U_{n \mu} = i (\lambda_{n} U_{n \mu} 
- U_{n \mu} \lambda_{n+\hat{\mu}} ) \ .
\label{symmetry-infinitesimal-lgt}
\end{alignat}
Here $\lambda_{n}$ is an element of the Lie algebra,
which can be expanded as
\begin{alignat}{3}
(\lambda_{n})_{jk} = \sum_a \lambda_{na} (t_a)_{jk} 
\label{lambda-expand-lgt}
\end{alignat}
in terms of 
the generators $t_a$ of ${\rm SU}(3)$ 
with real coefficients $\lambda_{na} \in \bbR$ .

When one complexifies the variables 
$U_{n \mu} \mapsto {\cal U}_{n \mu} \in  {\rm SL}(3,\bbC)$,
the symmetry of the action and the observables
naturally enhances to the ${\rm SL}(3,\bbC)$ gauge symmetry
that can be obtained by complexifying the original Lie group.
For instance, the plaquette action (\ref{plaquette-action}) becomes
\begin{alignat}{3}
S_{\rm plaquette}({\cal U}) = - \beta \sum_{n}  \sum_{\mu \neq \nu}
\tr ({\cal U}_{n\mu} {\cal U}_{n+ \hat{\mu},\nu} 
{\cal U}_{n+\hat{\nu},\mu}^{-1} {\cal U}_{n\nu}^{-1} )  \ ,
\label{plaquette-action-complexified}
\end{alignat}
which is invariant under
\begin{alignat}{3}
{\cal U}_{n \mu} ' = g_{n} {\cal U}_{n \mu} g_{n+\hat{\mu}}^{-1}
\label{symmetry-complexified-lgt}
\end{alignat}
with $g_{n} \in {\rm SL}(3,\bbC)$.
An infinitesimal transformation is given by
\begin{alignat}{3}
\delta {\cal U}_{n \mu} = i (\lambda_{n} {\cal U}_{n \mu} 
- {\cal U}_{n \mu} \lambda_{n+\hat{\mu}} ) \ .
\label{symmetry-infinitesimal-complexified-lgt}
\end{alignat}
Here $\lambda$ is an element of the Lie algebra
for the complexified Lie group, which can be expanded
as (\ref{lambda-expand-lgt}) but now with complex coefficients
$\lambda_{na} \in \bbC$.

The discretized version of the
complex Langevin equation 
(\ref{eq:complex-Langevin-lgt})
can be written as
\begin{alignat}{3}
{\cal U}_{n \mu}^{(\eta)} (t+\epsilon) = 
\exp \Big\{
i  \sum_a 
\Big( - \epsilon {\cal D}_{a n \mu} S({\cal U}) 
+ \sqrt{\epsilon} \eta_{a n \mu}(t) \Big)
\, t_a \Big\}
\,  {\cal U}_{n \mu}^{(\eta)} (t) \ .
\label{eq:Langevin-discretized2-complexified-lgt}
\end{alignat}
The gauge cooling \cite{Seiler:2012wz}
modifies the complex Langevin equation 
(\ref{eq:Langevin-discretized2-complexified-lgt}) into
\begin{alignat}{3}
\widetilde{\cal U}_{n \mu}^{(\eta)} (t) & = 
g_{n} \, 
{\cal U}_{n \mu}^{(\eta)} (t) \, 
g_{n+\hat{\mu}}^{-1}   \ ,  \nonumber \\
{\cal U}_{n \mu}^{(\eta)} (t+\epsilon) & = 
\exp \Big\{
i \sum_a 
\Big( - \epsilon{\cal D}_{a n \mu} S(\widetilde{\cal U}) 
+ \sqrt{\epsilon} \eta_{a n \mu}(t) \Big)
\, t_a \Big\}
\,  \widetilde{\cal U}_{n \mu}^{(\eta)} (t)  \ ,
\label{eq:Langevin-discretized2-complexified-cooled-lgt}
\end{alignat}
where $g_{n}$ is an element of the complexified Lie group.
The basic idea is to determine $g_{n}$
in such a way that the modified Langevin process 
(\ref{eq:Langevin-discretized2-complexified-cooled-lgt})
does not suffer from the problem of the original 
Langevin process (\ref{eq:Langevin-discretized2-complexified-lgt}).

For instance, if the excursions in the imaginary directions
are problematic,
one can introduce 
a positive semi-definite quantity \cite{Aarts:2008rr}
(We call it the ``norm'' in this paper.)
\begin{alignat}{3}
{\cal N} = \sum_{n \mu}
\tr ({\cal U}_{n\mu}^\dagger {\cal U}_{n\mu}  - {\bf 1}) \ ,  
\label{unitarity-norm}
\end{alignat}
which measures the distance from the unitary region,
and determine the transformation $g_{n}$ 
in (\ref{eq:Langevin-discretized2-complexified-cooled-lgt})
in such a way that
the norm ${\cal N}$ for ${\cal U}^{(\eta)} (t)$
is reduced by the transformation.
Typically the norm one tries to reduce is invariant under
transformations in the original Lie group but not under
transformations in the complexified Lie group
as in the case of (\ref{unitarity-norm}).
Below we demonstrate that the modification of the Langevin process
by the gauge cooling does not spoil the equivalence
to the path integral reviewed in the previous section.

Note that the gauge cooling is a completely deterministic procedure.
In particular, the transformation $g_{n}$ in
(\ref{eq:Langevin-discretized2-complexified-cooled-lgt})
is determined only by the configuration
${\cal U}^{(\eta)} (t)$
before cooling.
Therefore, for our purpose, 
it is more convenient to rewrite 
(\ref{eq:Langevin-discretized2-complexified-cooled-lgt})
into the form
\begin{alignat}{3}
{\cal U}_{n \mu}^{(\eta)} (t+\epsilon) & = 
g_{n} \, 
\exp \Big\{
i  \sum_a 
\Big( - \epsilon {\cal D}_{a n \mu} S({\cal U}) t_a 
+ \sqrt{\epsilon}
\eta_{a n \mu}(t) g_{n}^{-1} t_a g_{n} \Big) \Big\}
\,  {\cal U}_{n \mu}^{(\eta)} (t) 
g_{n+\hat{\mu}}^{-1} \ ,
\label{eq:Langevin-discretized2-complexified-cooled-tilde-lgt}
\end{alignat}
which involves ${\cal U}^{(\eta)} (t)$ only.\footnote{In practice,
one usually measures observables using the configuration
$\tilde{\cal U}^{(\eta)}(t)$ after cooling instead 
of ${\cal U}^{(\eta)}(t)$.
This does not cause any problem 
since ${\cal U}^{(\eta)}(t)$ and 
$\tilde{\cal U}^{(\eta)}(t)$ are
related to each other 
by the complexified symmetry 
transformation (\ref{symmetry-complexified-lgt}),
under which the observables are invariant.
}
Let us assume that 
$g_{n}$ is given as
\begin{alignat}{3}
g_{n}
& = 
\exp \Big\{  i \epsilon 
\lambda_{n} \Big({\cal U}^{(\eta)} (t) \Big)
 \Big\} \ , 
\label{lambda-def-lgt}
\end{alignat}
where 
$\lambda_{n} ({\cal U})$ is 
an element of the Lie algebra of 
${\rm SL}(N,\bbC)$.
For instance, one may use
\begin{alignat}{3}
\lambda_{n} ({\cal U})
 &=  
 i \alpha({\cal U})  \sum_{\mu} 
\Big\{
(  {\cal U}_{n\mu} {\cal U}_{n\mu}^\dagger 
- {\cal U}_{n-\hat{\mu} ,\mu}^\dagger  {\cal U}_{n-\hat{\mu} , \mu} 
) - (\mbox{trace part}) \Big\} \ , 
\label{eq:gradient-lgt}
\end{alignat}
which can be obtained by calculating the gradient of 
the norm (\ref{unitarity-norm}) with respect to
the ${\rm SL}(3,\bbC)$ gauge transformation of the configuration ${\cal U}$.
The real positive function 
$\alpha({\cal U})$, which is not necessarily holomorphic,
can be chosen to optimize the reduction of the norm.
(In practice, one can simply determine $g_{n}$ 
by using the steepest descent method
with respect to the norm (\ref{unitarity-norm}) \cite{Seiler:2012wz}.)
Note that $\lambda_{n} ({\cal U})$
is not a holomorphic function of ${\cal U}_{n \mu}$
in general as in (\ref{eq:gradient-lgt}).

Using (\ref{lambda-def-lgt})
in eq.~(\ref{eq:Langevin-discretized2-complexified-cooled-tilde-lgt})
and taking the $\epsilon \rightarrow 0$ limit,
we obtain the continuum complex Langevin equation 
for ${\cal U}^{(\eta)} (t)$ as
\begin{alignat}{3}
\dot{{\cal U}}_{n \mu}^{(\eta)} (t) =& 
i \sum_a 
\Big( - {\cal D}_{a n \mu} S({\cal U})
+ \eta_{a n \mu}(t) \Big)
\, t_a \, {\cal U}_{n \mu}^{(\eta)} (t)  \nonumber \\
&+ i \Bigl\{
\lambda_{n} \Big({\cal U}^{(\eta)} (t) \Big) \,
{\cal U}_{n \mu}^{(\eta)} (t) 
- {\cal U}_{n \mu} ^{(\eta)} (t) \, 
\lambda_{n+\hat{\mu}} 
\Big({\cal U}^{(\eta)} (t) \Big)
\Bigr\} 
\ ,
\label{eq:complex-Langevin-cooling-lgt}
\end{alignat}
where the effect of the gauge cooling is represented by the
last term on the right-hand side. 
Then we can easily find that the FP-like equation (\ref{FP-like-eq-lgt}) 
that $P({\cal U};t)$ 
satisfies is modified by the gauge cooling as
\begin{alignat}{3}
\frac{\del P}{\del t}
= & {\cal D}_{a n \mu}^{\rm (R)}
\left\{ 
{\rm Re} \Big( {\cal D}_{a n \mu} S ({\cal U}) - {\cal C}_{a n \mu} \Big)
+ N_{\rm R} {\cal D}_{a n \mu}^{\rm (R)} \right\} P 
\nonumber \\
& + {\cal D}_{a n \mu}^{\rm (I)}
\left\{ 
{\rm Im} \Big( {\cal D}_{a n \mu} S ({\cal U})- {\cal C}_{a n \mu} \Big)
+ N_{\rm I} {\cal D}_{a n \mu}^{\rm (I)} \right\} P \ ,
\label{FP-like-eq-cooling-lgt}
\\
{\cal C}_{a n \mu} =& \tr \left\{ t_a 
\Bigl(\lambda_{n}({\cal U})  - 
{\cal U}_{n \mu} 
\lambda_{n+\hat{\mu}}({\cal U})
\, {\cal U}_{n \mu}^{-1} \Bigr) \right\} \ .
\label{def-c-lgt}
\end{alignat}
This modifies the differential operator $L$ 
in eq.~(\ref{L-expression-lgt}) into 
\begin{alignat}{3}
L =& 
\left\{ 
- {\rm Re} \Big( {\cal D}_{a n \mu} S ({\cal U})
- {\cal C}_{a n \mu} \Big)
+ N_{\rm R} {\cal D}_{a n \mu}^{\rm (R)}
\right\}
{\cal D}_{a n \mu}^{\rm (R)}  
\nonumber \\
& +
\left\{ 
- {\rm Im} \Big( {\cal D}_{a n \mu} S ({\cal U})
- {\cal C}_{a n \mu} \Big)
+ N_{\rm I} {\cal D}_{a n \mu}^{\rm (I)}
\right\}
{\cal D}_{a n \mu}^{\rm (I)}  \ . 
\label{L-expression-mod-lgt}
\end{alignat}
Acting this modified $L$ 
on a holomorphic function $f({\cal U})$,
we obtain
\begin{alignat}{3}
L f({\cal U}) &= 
\left\{ - {\cal D}_{a n \mu} S ({\cal U}) + {\cal C}_{a n \mu} 
+ (N_{\rm R}- N_{\rm I} ) {\cal D}_{a n \mu} 
\right\}
{\cal D}_{a n \mu}   f({\cal U}) \nonumber \\
& = 
\tilde{L} f({\cal U}) + {\cal C}_{a n \mu}  {\cal D}_{a n \mu} 
f({\cal U}) 
 \ ,
\label{Lf-modified-lgt}
\end{alignat}
where $\tilde{L} $ is defined by (\ref{Lftilde-lgt}).
The extra term compared with
(\ref{Lf-lgt})
represents the change of $f({\cal U})$
under an infinitesimal ${\rm SL}(3,\bbC)$ gauge transformation.
Note that 
the time-evolved observables ${\cal O}({\cal U};t)$ 
defined by (\ref{Ot-def-lgt})
remain to be invariant under the 
${\rm SL}(3,\bbC)$ gauge transformation
as long as the action $S({\cal U})$
and the original observables ${\cal O}({\cal U})$ are invariant.
Therefore, the $\tilde{L}$ in the first term of 
(\ref{del-F-lgt}) can be replaced by the modified $L$.
Hence (\ref{del-F-lgt}) vanishes
if one can perform integration by parts
for the modified $L$ and $P$.
In that case, the crucial identity (\ref{P-rho-rel-lgt}) 
holds for the modified $P$ with the same $\rho$.
Thus we have shown explicitly that the gauge cooling
provides a possibility to improve 
the property of the probability distribution $P({\cal U};t)$ 
so that (\ref{P-rho-rel-lgt}) holds, without affecting
the FP equation (\ref{FPeq-complex-lgt}) for $\rho(U;t)$.

\section{Summary and discussions} 
\label{sec:conclusion}

In this paper we have provided a rigorous justification
of the CLM with the gauge cooling procedure.
As we have reviewed in detail,
the CLM relies crucially
on the relation between
the probability distribution $P$ associated with the
complex Langevin process and 
the complex weight $\rho$ associated with the original
path integral problem.
This relation holds if and only if
the probability distribution $P$ satisfies the following two properties.
One is that it 
has a rapid fall-off in the imaginary directions
in the complexified configuration space \cite{Aarts:2009uq,Aarts:2011ax}.
The other is that  
the distribution is such that
complexified configurations which make the
drift term singular are strongly suppressed \cite{Nishimura:2015pba}.
Since the gauge cooling modifies
the probability distribution $P$,
one may hope to make it satisfy the above two properties
by appropriately choosing the complexified symmetry transformation
to be used in the cooling procedure.
What we have shown in this paper is that
the modification of the probability distribution $P$
due to the gauge cooling does not alter the FP equation
that the complex weight $\rho$ obeys
if the relation between $P$ and $\rho$ holds at all.


For a long time it has been thought that the convergence of
the FP equation for the complex weight $\rho$ is not guaranteed
unlike in the real action case.
However, once the relation between 
the probability distribution $P$ and 
the complex weight $\rho$ is established,
one may argue \cite{Nishimura:2015pba}
that the unique convergence of 
the probability distribution $P$ 
already implies the unique convergence of $\rho$
to the desired complex weight $\ee^{-S}$.
Therefore, if one can satisfy the above two properties
of the probability distribution $P$ by 
using the gauge cooling appropriately,
the CLM is guaranteed to give the correct results.

While the gauge cooling certainly enlarges the range of 
applicability of the CLM, it remains to be seen how powerful
it is in studying various interesting systems with complex actions.
In this regard, our results for the Random Matrix Theory using
the gauge cooling with a new type of norm \cite{NNS}
look very promising.

\section*{Acknowledgements}

The authors would like to 
thank J.~Bloch, K.~Fukushima and D.~Sexty for valuable discussions.
We are also grateful to E.~Seiler for correspondence 
on the first version of this paper.
K.~N.\ was supported by JSPS Grants-in-Aid for Scientific Research (Kakenhi)
Grants No. 00586901,  MEXT SPIRE and JICFuS.
The work of J.~N.\ was supported in part by Grant-in-Aid 
for Scientific Research (No.\ 23244057)
from Japan Society for the Promotion of Science.

\appendix

\section{Derivation of eq.~(\ref{FP-like-eq-lgt})}
\label{sec:derive-FP-like-eq-lgt}

In this appendix, we derive 
the Fokker-Planck like equation (\ref{FP-like-eq-lgt})
to make this paper self-contained.
Here we deal with the continuous $t$ for simplicity, but
one can make a similar analysis with discretized $t$ 
as we have done in section \ref{sec:discretized-Langevin}.

Let us consider a test function $f({\cal U})$ and its expectation value
\begin{alignat}{3}
\left\langle f \Big({\cal U}^{(\eta)}(t) \Big) \right\rangle_{\eta}
=  \int d  {\cal U}  \, 
f({\cal U})
P({\cal U};t) \ .
\label{test-fn-lgt}
\end{alignat}
Taking the derivative with respect to the fictitious time $t$, we get
\begin{alignat}{3}
\frac{d}{dt} \left\langle f \Big({\cal U}^{(\eta)}(t) \Big) 
\right\rangle_{\eta}
=&   \int d  {\cal U}   \, 
f({\cal U})
\frac{d}{dt}  P({\cal U};t) \ .
\label{test-fn-lgt-deriv}
\end{alignat}
The left-hand side can be evaluated as follows.
\begin{alignat}{3}
 \frac{d}{dt} \left\langle f 
\Big({\cal U}^{(\eta)}(t) \Big) \right\rangle_{\eta}
=&
\left\langle 
\Big\{ - {\rm Re} 
\Big({\cal D}_{a n \mu} S({\cal U}^{(\eta)}(t)) \Big)
+ \eta_{a n \mu}^{\rm (R)}(t)  \Big\} \, 
{\cal D}_{a n \mu}^{\rm (R)}
 f  \Big({\cal U}^{(\eta)}(t) \Big) 
\right\rangle_{\eta}
\nonumber \\
&+
\left\langle 
\Big\{ - {\rm Im} \Big( {\cal D}_{a n \mu} S({\cal U}^{(\eta)}(t)) \Big)
+ \eta_{a n \mu}^{\rm (I)}(t) \Big\} \, 
{\cal D}_{a n \mu} ^{\rm (I)}
f \Big({\cal U}^{(\eta)}(t)\Big) 
\right\rangle_{\eta}  \ .
\label{test-fn-lgt-lhs}
\end{alignat}
Here we use the following formula. 
(See ref.~\cite{Damgaard:1987rr}, for instance.)
\begin{alignat}{3}
 \left\langle 
g \Big({\cal U}^{(\eta)}(t) \Big) \eta_{a n \mu}^{\rm (R)}(t) 
\right\rangle_{\eta} 
&=
\left\langle 
2 N_{\rm R} 
\frac{\delta}{\delta \eta_{a n \mu}^{\rm (R)}(t)} 
g \Big({\cal U}^{(\eta)}(t)\Big) 
\right\rangle_{\eta}
 =
\Big\langle 
  N_{\rm R} {\cal D}_{a n \mu}^{\rm (R)}
g \Big({\cal U}^{(\eta)}(t)  \Big) 
\Big\rangle_{\eta} \ , 
\label{test-fn-lgt-lhs2}
\\
 \left\langle 
g \Big({\cal U}^{(\eta)}(t) \Big) \eta_{a n \mu}^{\rm (I)}(t) 
\right\rangle_{\eta} 
&=
\left\langle 
2 N_{\rm I} 
\frac{\delta}{\delta \eta_{a n \mu}^{\rm (I)}(t)} 
g \Big({\cal U}^{(\eta)}(t)\Big) 
\right\rangle_{\eta}
 =
\Big\langle 
  N_{\rm I} {\cal D}_{a n \mu}^{\rm (I)}
g \Big({\cal U}^{(\eta)}(t)  \Big) 
\Big\rangle_{\eta} \ .
\label{test-fn-lgt-lhs3}
\end{alignat}
Using (\ref{test-fn-lgt-lhs2}) and (\ref{test-fn-lgt-lhs3})
in (\ref{test-fn-lgt-lhs}), we get
\begin{alignat}{3}
& \frac{d}{dt} 
\left\langle f \Big({\cal U}^{(\eta)}(t) \Big) \right\rangle_{\eta}
\nonumber
\\
= & - \left\langle 
 {\rm Re} \Big( {\cal D}_{a n \mu} S({\cal U}^{(\eta)}(t)) 
\Big) \, 
{\cal D}_{a n \mu}^{\rm (R)} f \Big({\cal U}^{(\eta)}(t) \Big) 
\right\rangle_{\eta}
+ 
\left\langle 
  N_{\rm R} {\cal D}_{a n \mu}^{\rm (R)}  
{\cal D}_{a n \mu}^{\rm (R)}  
 f \Big({\cal U}^{(\eta)}(t) \Big) 
\right\rangle_{\eta}
\nonumber \\
&
- \left\langle 
 {\rm Im} \Big( {\cal D}_{a n \mu} S({\cal U}^{(\eta)}(t)) 
\Big) \, 
{\cal D}_{a n \mu}^{\rm (I)} f \Big({\cal U}^{(\eta)}(t) \Big) 
\right\rangle_{\eta}
+ 
\left\langle 
  N_{\rm I} {\cal D}_{a n \mu}^{\rm (I)}  
{\cal D}_{a n \mu}^{\rm (I)}  
 f \Big({\cal U}^{(\eta)}(t) \Big) 
\right\rangle_{\eta}
\nonumber \\
= &
\int d  {\cal U}   \, P({\cal U};t) 
\Big[
- 
{\rm Re} \Big( {\cal D}_{a n \mu} S({\cal U}) \Big) 
{\cal D}_{a n \mu}^{\rm (R)}   f ({\cal U}) 
+  N_{\rm R} {\cal D}_{a n \mu}^{\rm (R)} 
{\cal D}_{a n \mu}^{\rm (R)} f ({\cal U})
\nonumber \\
& \quad \quad \quad \quad \quad \quad \quad 
 - 
{\rm Im} \Big( {\cal D}_{a n \mu} S({\cal U}) \Big) 
{\cal D}_{a n \mu}^{\rm (I)}   f ({\cal U}) 
+
 N_{\rm I} {\cal D}_{a n \mu}^{\rm (I)} 
{\cal D}_{a n \mu}^{\rm (I)} f ({\cal U})
\Big]
\nonumber \\
=&
\int d  {\cal U}   \, 
f({\cal U})
\Big[
 {\cal D}_{a n \mu}^{\rm (R)} 
\left\{ 
{\rm Re} \Big({\cal D}_{a n \mu} S ({\cal U}) \Big)
+ N_{\rm R} {\cal D}_{a n \mu} ^{\rm (R)} 
\right\} P 
\nonumber \\
& 
\quad\quad\quad\quad\quad\quad + 
 {\cal D}_{a n \mu}^{\rm (I)} 
\left\{ 
{\rm Im} \Big({\cal D}_{a n \mu} S ({\cal U}) \Big)
+ N_{\rm I} {\cal D}_{a n \mu} ^{\rm (I)} 
\right\} P
\Big]  \ .
\label{test-fn-lgt-lhs-fin}
\end{alignat}
Plugging this expression in (\ref{test-fn-lgt-deriv}), and using
the fact that (\ref{test-fn-lgt-deriv}) should hold for 
an arbitrary $f({\cal U})$,
we get eq.~(\ref{FP-like-eq-lgt}).


\end{document}